\newcommand{\bear}{\begin{eqnarray}}
\newcommand{\eear}{\end{eqnarray}}
\documentclass[12pt]{article}
\pdfoutput=1
\usepackage{threeparttable}
\usepackage{graphicx}
\usepackage{a4wide,amsmath,amssymb,setspace}
\usepackage[figuresright]{rotating}
\begin{document}
\title{\bf{OBSERVATION OF CHAOTIC  BEATS IN A  DRIVEN 
MEMRISTIVE CHUA'S CIRCUIT}}
\author{A.~ISHAQ AHAMED$^{\ast}$, K.~SRINIVASAN$^{\dagger}$, K.~MURALI$^{\ddagger}$, M.~LAKSHMANAN$^{\dagger}$\\
$^{\ast}$Department of Physics, Jamal Mohamed College,\\
Tiruchirappalli - 620 020, India\\ 
$^{\dagger}$Centre for Nonlinear Dynamics, School of Physics, \\
Bharathidasan University, Tiruchirappalli - 620 024, India\\ 
$^{\ddagger}$Department of Physics, Anna University,\\
Chennai -600 025, India}
\date{\today}
\maketitle
\noindent
Email: lakshman@cnld.bdu.ac.in \\
Running Title : Beats in Memristive Driven Chua's Circuit  \\
Corresponding author : M. Lakshmanan  

\begin{abstract} 
In this paper, a time varying resistive circuit realising the action of an active three segment piecewise linear flux controlled memristor is proposed. Using this as the nonlinearity, a driven Chua's circuit is implemented. The phenomenon of chaotic beats in this circuit is observed for  a suitable choice of parameters. The memristor acts as a chaotically time varying resistor (CTVR), switching between a less conductive OFF state and a more conductive ON state. This chaotic switching is governed by the dynamics of the driven Chua's circuit of which the memristor is an integral part. The occurrence of beats is essentially due to the interaction of the memristor aided self oscillations of the circuit and the external driving sinusoidal forcing. Upon slight tuning/detuning of the frequencies of the memristor switching and that of the external force, constructive and destructive interferences occur leading to revivals and collapses in amplitudes of the circuit variables, which we refer as chaotic beats. Numerical simulations and Multisim modelling as well as statistical analyses have been carried out to observe as well as to understand and verify the mechanism leading to chaotic beats.\\ \\

\end{abstract}

\section{\label{sec:level1}Introduction} 

The phenomenon of beats arises due to the superposition of two periodic excitations of slightly different frequencies. This superposition results in a cyclic behaviour arising due to the constructive and destructive interferences between these two excitations. The resultant waveform is characterized by a frequency that equals the average of the frequencies of the two waves, whereas its amplitude is modulated by an envelope, whose frequency known as the {\it{beat frequency}} is the difference between the frequencies of the two waves. This phenomenon has been widely studied in linear systems. However recently, attention has been drawn to chaotic and hyper-chaotic beats in nonlinear systems. Chaotic or hyper-chaotic beats arise due to the irregular collapses and revivals in amplitudes of the variables of nonlinear systems as a result of the interaction between two different excitations (either self oscillations or external driving forces) in these systems. Though they can be identified by visual inspection, the presence of positive Lyapunov exponent(s) is considered as the true characterizing feature of chaotic or hyper-chaotic beats.   Chaotic and hyper-chaotic beats were identified for the first time in a system of coupled Kerr oscillators and coupled Duffing oscillators with small nonlinearities and strong external pumping [Grygiel $\&$ Szlachetka, 2002]. Chaotic beats were also reported in coupled non-autonomous Chua's circuits [Cafagna $\&$ Grassi, 2004]. Weakly chaotic and hyper-chaotic beats were reported in individual and coupled nonlinear optical subsystems, respectively, describing second harmonic generation (SHG) of light [\'{S}liwa {\it{et al}}., 2008]. In all these coupled systems [Grygiel $\&$ Szlachetka, 2002; Cafagna $\&$ Grassi, 2004; \'{S}liwa {\it{et al}}., 2008],  the occurrence of beats is attributed to the interaction between the self oscillations or driven oscillations of each of the coupled subsystems. Using extensive Pspice simulations, the occurrence of beats in individual driven Chua's circuit with two external excitations has also been reported [Cafagna $\&$ Grassi, 2006a; 2006b]. Chaotic beats have also been found to occur in individual Murali- Lakshmanan-Chua (MLC) circuit by the same authors [Cafagna $\&$ Grassi, 2005]. In these cases of single systems [Cafagna $\&$ Grassi, 2005; 2006a; 2006b], two different external driving forces with slightly differing frequencies were the cause for the occurrence of beats. 

On the other hand, from a different perspective, Leon Chua in the year 1971 introduced a fourth circuit element, namely the memristor [Chua, 1971; Chua $\&$ Kang, 1976], purely from theoretical arguments, and pointed out the interesting possibilities of including it in electronic circuits. Almost four decades later, Strutkov et al.  of  Hewlett-Packard Laboratories designed the first approximate physical example of a memristor [Strutkov, 2008] in a solid-state nano-scale system in which electronic and ionic transports are coupled under an external bias voltage. As the  $i-v$ characteristic of  memristors is inherently non-linear and is unique in the sense that no combination of nonlinear resistive, capacitive and inductive components can duplicate their circuit properties, memristors have generated considerable excitement among circuit theorists. Further they provide new circuit functions such as electronic resistance switching at extremely high two-terminal device densities [Strutkov, 2008]. Very recently Itoh and Chua [2009], have investigated the dynamics of nonlinear electronic circuits of different orders, including autonomous Chua's oscillator and canonical Chua's oscillator by replacing a Chua's diode - a well studied and familiar nonlinear device - by an {\it{active}} or a {\it{passive}} memristor. Hence it will be of high current interest to study the effect of replacing the Chua's diode by a memristor in a driven Chua's circuit [Murali $\&$ Lakshmanan, 1990; 1991], and investigate the new dynamical phenomena which arise from the corresponding circuit.

In the present paper, we report the occurrence of chaotic beats in the driven Chua's circuit using a flux controlled active memristor as its nonlinearity. The importance of this circuit is that it employs just a single driving force and makes use of the chaotically time varying resistive property and the switching characteristic of the memristor to generate beats. The plan of the paper is as follows. In Sec. 2 memristors, their characteristic and their behaviours are briefly described. In Sec. 3.1 the standard driven Chua's circuit is introduced while in Sec. 3.2 the driven Chua's circuit using a memristor as the nonlinear element is described. In Sec. 4 the chaotic beats phenomenon observed numerically in this circuit is shown with the aid of time series plots, phase portraits and Poincar\'{e} maps, Lyapunov exponents and power spectral calculations. In Sec. 5 an experimentally realisable circuit for the memristor is proposed. Using Multisim modelling, a driven Chua's circuit is constructed and the chaotic beats observed. With the help of the grapher facility of the software package, time series plots, phase portrait, power spectra  which are qualitatively equivalent to the numerical observations are presented. In Sec. 6 the mechanism for the occurrence of chaotic beats is discussed, while Sec. 7 presents a brief conclusion. 

\section{\label{mem}Memristors}

From a circuit theoretic point of view, Leon Chua [1971] noted that there could possibly be six mathematical relations connecting pairs of the four fundamental variables, namely charge $q$, current $i$, flux $ \phi$ and voltage $v$. One of these relations (the charge is time integral of current) is determined from the definitions of the two variables. Another relation (the flux is the time integral of the electromotive force or voltage) is given by Faraday's law of  induction. Three other relations can be constructed from the axiomatic definitions of the three classical two terminal circuit elements, namely, the resistor (defined by the relationship between $v$ and $i$), the inductor (defined by the relationship between $\phi$ and $i$ ) and the capacitor (defined by the relationship between $q$ and$v$). One more relationship between $\phi$  and $q$ remains undefined. From logical and axiomatic points of view, as well as for the sake of completeness, the necessity for the existence of a fourth basic two terminal circuit element was postulated by Chua [1971]. This {\it{missing element}} purported to have a functional relationship connecting charge $q$ and flux $\phi$ is named as $\it{memristor}$. This hypothetical element is endowed with both a resistive property and a memory like nature and hence it is named so. In principle, memristors with almost any $\phi-q$ characteristic can be synthesised [Chua, 1971]. Although realized, albeit only recently, and that too in the nano-scale range, memristors have been successfully used as a conceptual tool for analysis in signal processing and nonlinear circuits, see for example [Tour $\&$ He, 2008]. In particular their nonlinear characteristic provides designers with new circuit functions like electronic resistance switching at extremely high two terminal densities [Strutkov {\it{et al}}., 2008]. 

\section{\label{Dyn}Dynamics of the Driven Memristive Chua's Circuit}

We will now study the dynamics of the driven memristive Chua's circuit to bring out the concept
of chaotic beats in it. For this purpose we shall first introduce the standard driven Chua's circuit.

\subsection{\label{Driv_Chua}Driven Chua's Circuit}

The driven Chua's circuit is a fourth order non-autonomous circuit first introduced by Murali 
and Lakshmanan in the year 1990 [Murali $\&$ Lakshmanan, 1990]. It is found to exhibit a large variety of
bifurcations such as period adding,  quasi-periodicity, intermittency, equal periodic bifurcations,
re-emergence of double hook and double scroll attractors, hysteresis and coexistence of multiple 
attractors, besides the standard bifurcations . Its dynamics in the environment of a sinusoidal 
excitation has been extensively  studied in a series of works [Murali $\&$ Lakshmanan, 1990; 1991; 1992a; 1992b; 1993]. Due to its simplicity and the rich  content of  nonlinear dynamical phenomena, the driven Chua's circuit  continues to evoke renewed interest by researchers in the 
field of non-linear electronics [Anishchenko {\it{et al}}., 1993; Zhu $\&$ Liu 1997; Elwakil 2002; Srinivasan {\it{et al}}., 2009].
\begin{figure}
\centering
\includegraphics[width=0.8\columnwidth]{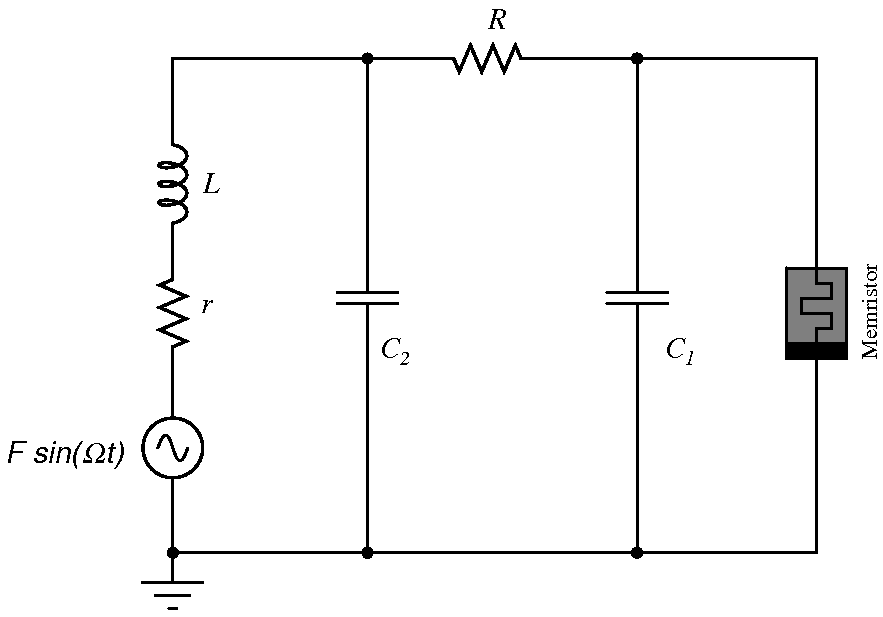}
\caption{Driven memristive Chua's circuit}
\end{figure}
\begin{figure}
\centering
\includegraphics[width=1.0\columnwidth]{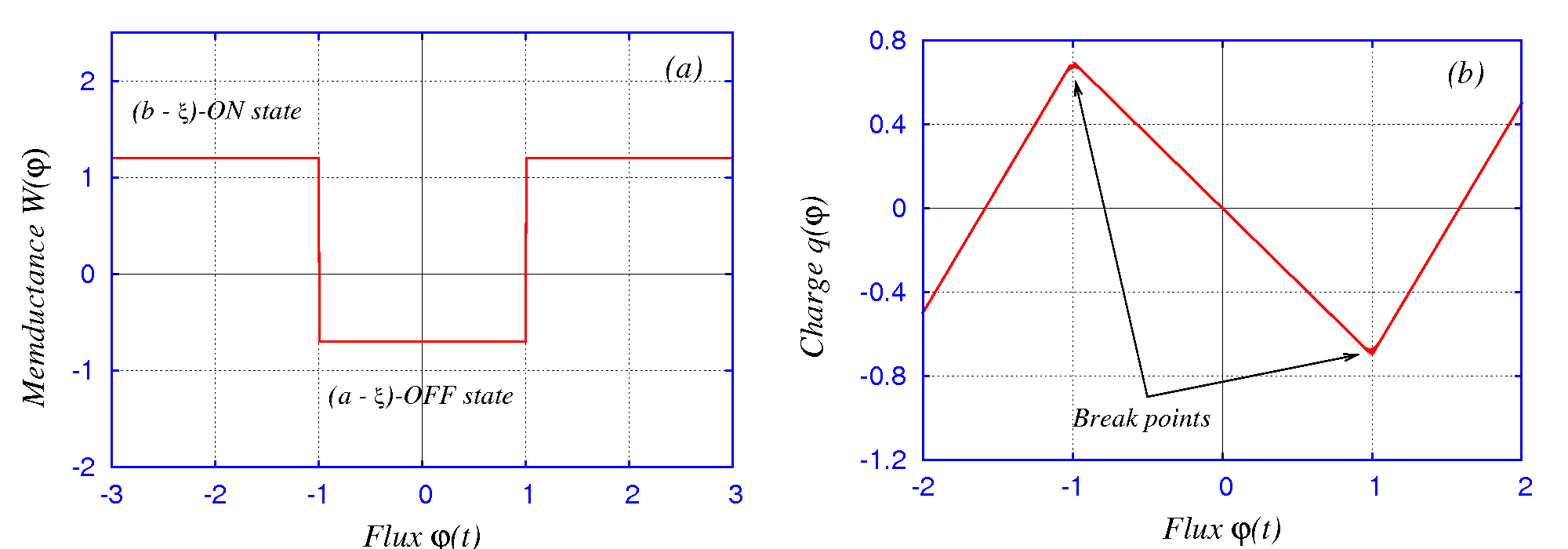}
\caption{(a) The variation of the memductance $W(\phi)$ as a function of the flux $\phi(t)$ across the memristor  and (b) the corresponding characteristic curve of the memristor in the $(\phi-q)$ plane.}
\end{figure}
\subsection{\label{Driv_Mem}Driven Memristive Chua's Circuit}

A large number of autonomous nonlinear electronic circuits using memristors as active nonlinear elements have been studied very recently by Itoh and Leon Chua [Itoh $\&$ Chua, 2008] and the conditions for the occurrence of chaos in them were mathematically analysed and numerically simulated.  Muthuswamy has  proposed a memristor having a cubic nonlinearity [Muthuswamy, 2009a; 2009b]. Using it a simple three element autonomous Chua's circuit was designed and its chaotic behaviour studied using MATLAB simulations, and also real time experiments. In this paper, we modify the driven Chua's circuit [Murali $\&$ Lakshmanan, 1990] mentioned above by removing an inductance and replacing the Chua's diode with a three segment {\it{piecewise-linear}} flux controlled active memristor [Itoh $\&$ Chua, 2008] as its nonlinearity. The circuit is given in Fig. 1.
\begin{figure}
\centering
\includegraphics[width=0.55\columnwidth=2]{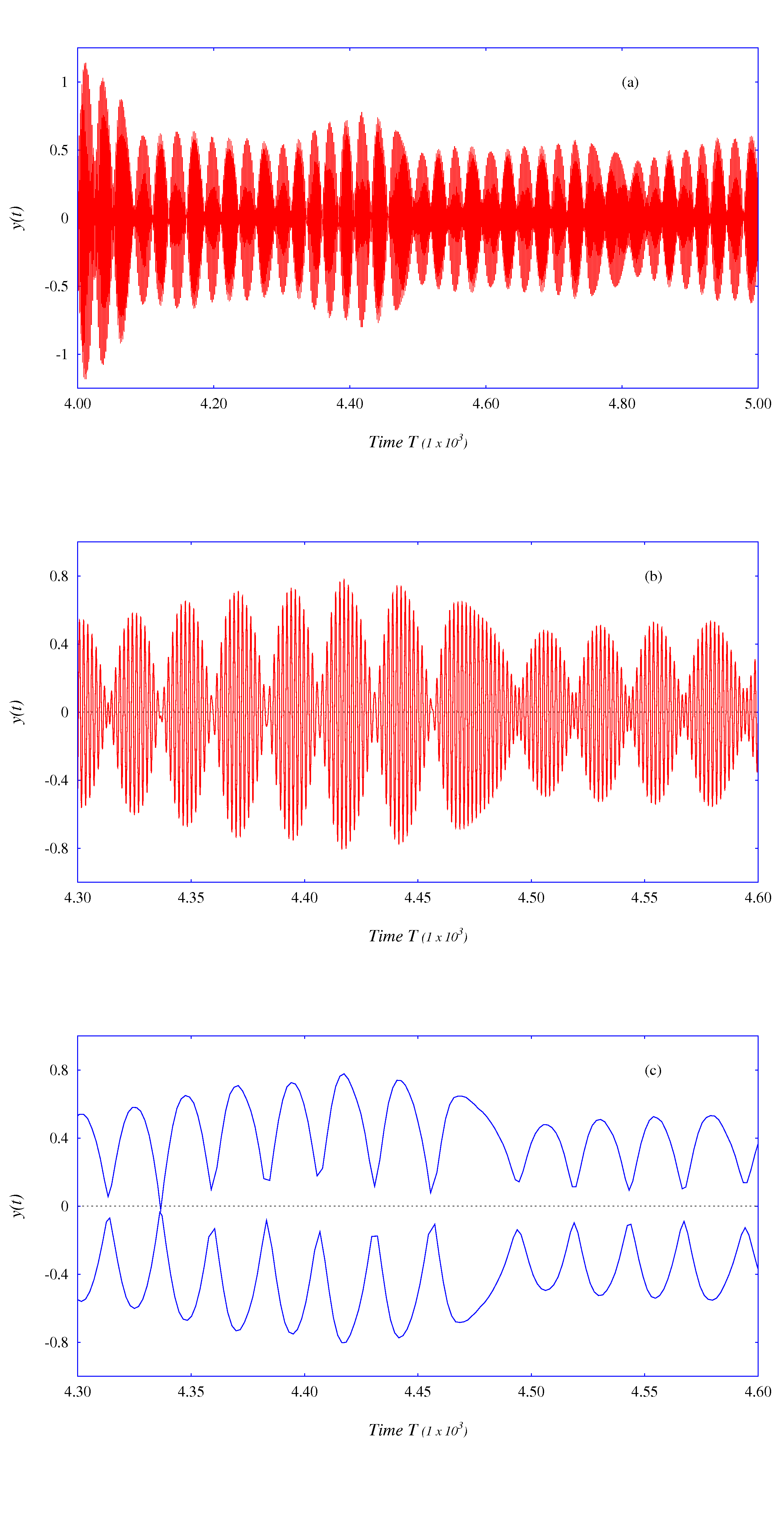}
\caption{Chaotic beats (Numerical simulation):  (a) A section of the time series of the variable y(t) showing the presence of amplitude modulation and chaotic beats. (b) An enlarged portion of the same portraying clearly the presence of a central frequency $\omega_c$ and variation in amplitudes. (c) The envelope of the time series given in (b) (obtained by Poincar\'{e} mapping technique), which is varying at a smaller frequency namely the beat frequency $\omega_b$.}
\end{figure}

An active flux controlled memristor  has the charge through it as some function of the flux across it. The functional relationship between the flux $\phi$ and charge $q$ is assumed to be the same expression as given in [Itoh $\&$ Chua, 2008], namely 

\begin{equation}
q(\phi)  =  (b-\xi)\phi + 0.5(a-b)[(|\phi+1|)-(|\phi-1|)],
\end{equation}
where a and b are the slopes in the inner and outer segments of the characteristic curve and $\xi$ is a negative conductance which is connected in parallel to the flux controlled memristor to make it satisfy the condition for activity.

The memristor employed here is an active device in the sense that it does not draw power from the circuit of which it is a part of, but in fact it supplies power to it. It satisfies the criterion for activity [Chua, 1971], namely
\begin{equation}
\int_{\tau_0}^{\tau}pd\tau   \leq  0,
\end{equation}
where
$p(t)  = W(\phi(t))i(t)^2  $is the instantaneous power dissipated across the memristor and $ W(\phi)$is the memductance of the memristor. The latter is so called because it has the units of conductance $(\mho)$. It is given by

\begin{equation}
W(\phi) = \frac{dq(\phi)}{d\phi} = \left \{\begin{array}{ll}
										(a-\xi), ~~~ | \phi  | \leq 1  \\
										(b-\xi), ~~~ | \phi  | >1
										\end{array}
								\right.
\end{equation}
where $a,b,\xi $ are as defined earlier.

The memductance of the memristor takes on two particular values as shown in Fig. 2(a), depending on the value of the flux across it. The higher memductance value $(b-\xi)$ can be referred as the ON state and the lesser memductance value $(a-\xi)$ can be referred as the OFF state. Obviously, as the flux across the memristor changes, the memristor switches or toggles between these two states. This switching action can be made use of profitably in utilising the memristor as a desirable element in modelling of nonlinear circuits. The characteristic curve in the $\phi-q$ plane which causes this memristive switching is shown in Fig. 2(b).

Following Itoh $\&$ Chua [2008], the circuit equations for the memristive driven Chua's circuit given by Fig. 1, are written as 

\begin{eqnarray}
C_1\frac{dv_1}{dt}  & = & \frac{v_2 - v_1}{R} - W(\phi)v_1,  \nonumber \\ 
C_2\frac{dv_2}{dt}  & = & \frac{v_1 - v_2}{R} - i, \nonumber\\
L \frac{di}{dt}  & = &  v_2 -ri +F\sin(\Omega t), \nonumber\\
\frac{d\phi}{dt}  & = & v_1.
\end{eqnarray}
Normalizing equations (4) using the same rescaling parameters as those used in [Itoh $\&$ Chua, 2008], namely
\begin{eqnarray}
x = v_1, y = v_2, z = -i, w = \phi, \nonumber \\
\alpha = \frac{1}{C_1}, \beta = \frac{1}{L}, \gamma = G, C_2 = 1, R = 1, f = \frac{F}{L}, \omega = \Omega,
\end{eqnarray}
 we have
\begin{eqnarray}
\dot{x}  & = & \alpha(y-x-W(\phi)x), \nonumber \\ 
\dot{y}  & = & x - y + z,\nonumber \\ 
\dot{z}  & = & -\beta y -\gamma z -f \sin(\omega t),\nonumber \\ 
 \dot{w}  & = & x.
\end{eqnarray}
Here dot stands for differentiation with respect to t. 

The piecewise linear function describing the memristor then becomes
\begin{equation}
W(w) = \frac{dq(w)}{dw} = \left \{\begin{array}{ll}
										(a-\xi), ~~~ | w  | \leq 1 \\
										(b-\xi), ~~~ | w  | > 1
										\end{array}
								\right.
								a,b,\xi > 0
\end{equation}
where $ a,b,\xi $ are now the normalized values of the quantities mentioned earlier.

\section{\label{Obsv}Observation of Chaotic Beats (Numerical Analysis)}

The above circuit, represented by the normalized equations (6), though simple in nature, generates beats due to the slight tuning / detuning of frequencies of the external driving excitation and the switching of the memristor between the ON and OFF memristive states. If we set $ \alpha = 10, \beta = 15.5, \gamma = 0.35, a = 2.0,b = 3.9, \xi = 2.7$ and change the force $f$ and frequency $\omega$  of the external signal suitably in equations (6), chaotic beats are found to occur in this circuit. For example for $f=0.75$ and $\omega  = 3.5$, chaotic beats are observed. This can be seen clearly in Figs. 3. While Fig. 3(a) shows a sampled stretch of the time series for the variable $y(t)$ depicting chaotic modulation in amplitudes, Fig. 3(b) shows an enlarged portion of the same illustrating the presence of a central fundamental frequency component, which we call as  $\omega_c$ .  In Fig. 3(c), the envelope of the amplitudes of the same variable $y(t)$ for the same stretch as in Fig. 3(b), isolated by Poincar\'{e} mapping technique, is captured. It shows that this variation of the amplitudes occurs at a smaller frequency, which we will call as $\omega_b$.
\begin{figure}
\includegraphics[width=1.0\columnwidth=4]{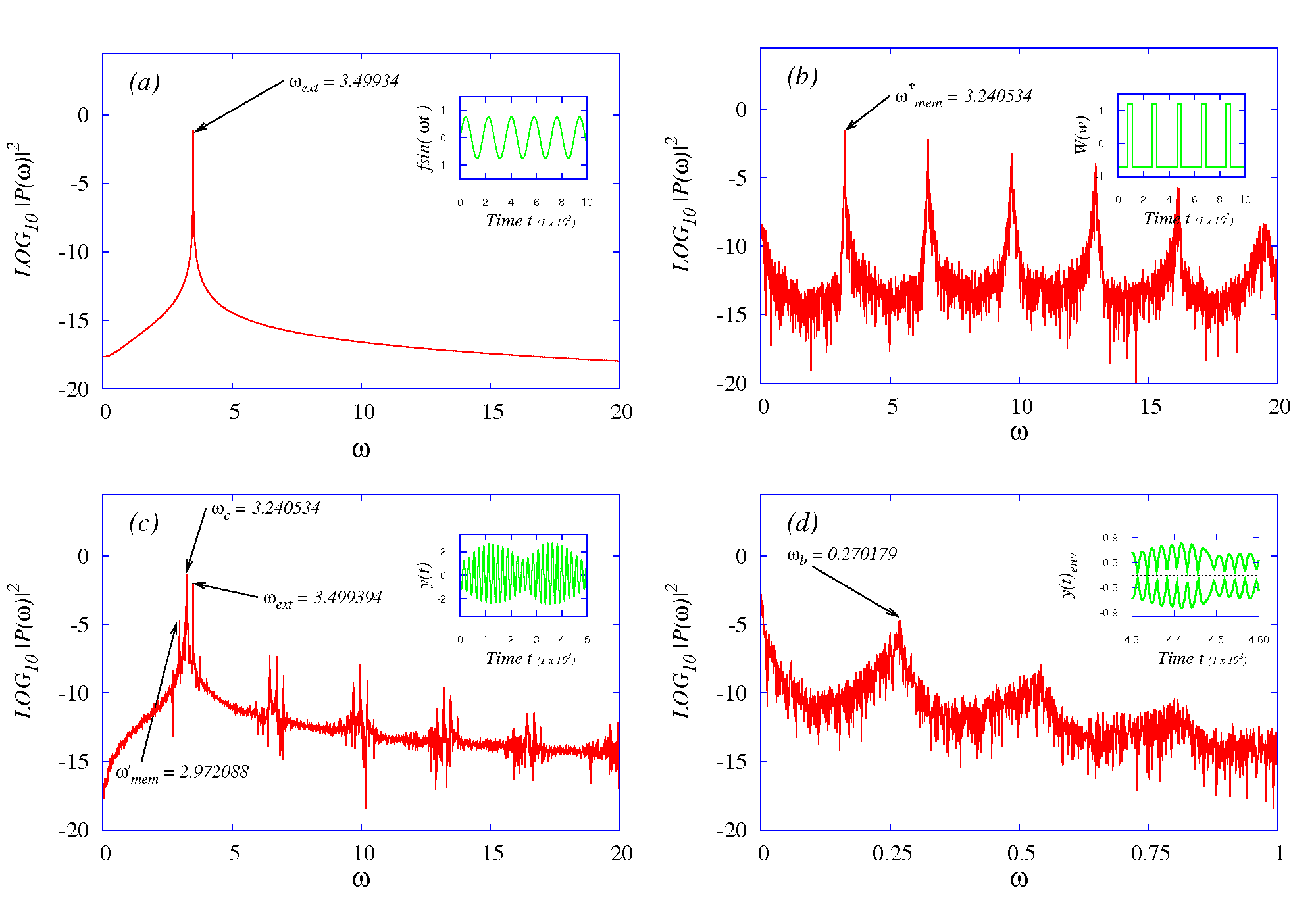}
\caption{The power spectrum (from numerical data) of (a) the external sinusoidal driving,(b) the variable $y(t)$ of the memristive Chua's oscillator when no external force is added, showing the characterisitc switching frequency $\omega^{*}_{mem}$ of the memristor,(c) the variable $y(t)$ of the driven memristive Chua's oscillator, showing the various frequency components $\omega^{*}_{mem}, ~\omega'_{mem}$ and  $\omega_c$ present in the chaotically modulated signal and(d)  the envelope showing the beat frequency ~$\omega_b$. The various signals whose power spectra are obtained are shown in the respective insets.}
\end{figure}

To find the frequencies of the circuit variables as well as to determine the frequency with which the memristor changes its ON memductive state to OFF state and vice versa, the power spectra are obtained as shown in Figs. 4, for four different cases, namely (i) for the external driving sinusoidal signal, (ii) for the memristor switching when driven by the self oscillations of the circuit, (iii) for the normalized voltage $y(t)$ of the driven Memristive Chua's oscillator  and (iv) for the envelope of the chaotically modulated signal. The various signals whose power spectra are obtained are shown as insets in Figs. 4.
\begin{figure}
\centering
\includegraphics[width=0.80\columnwidth]{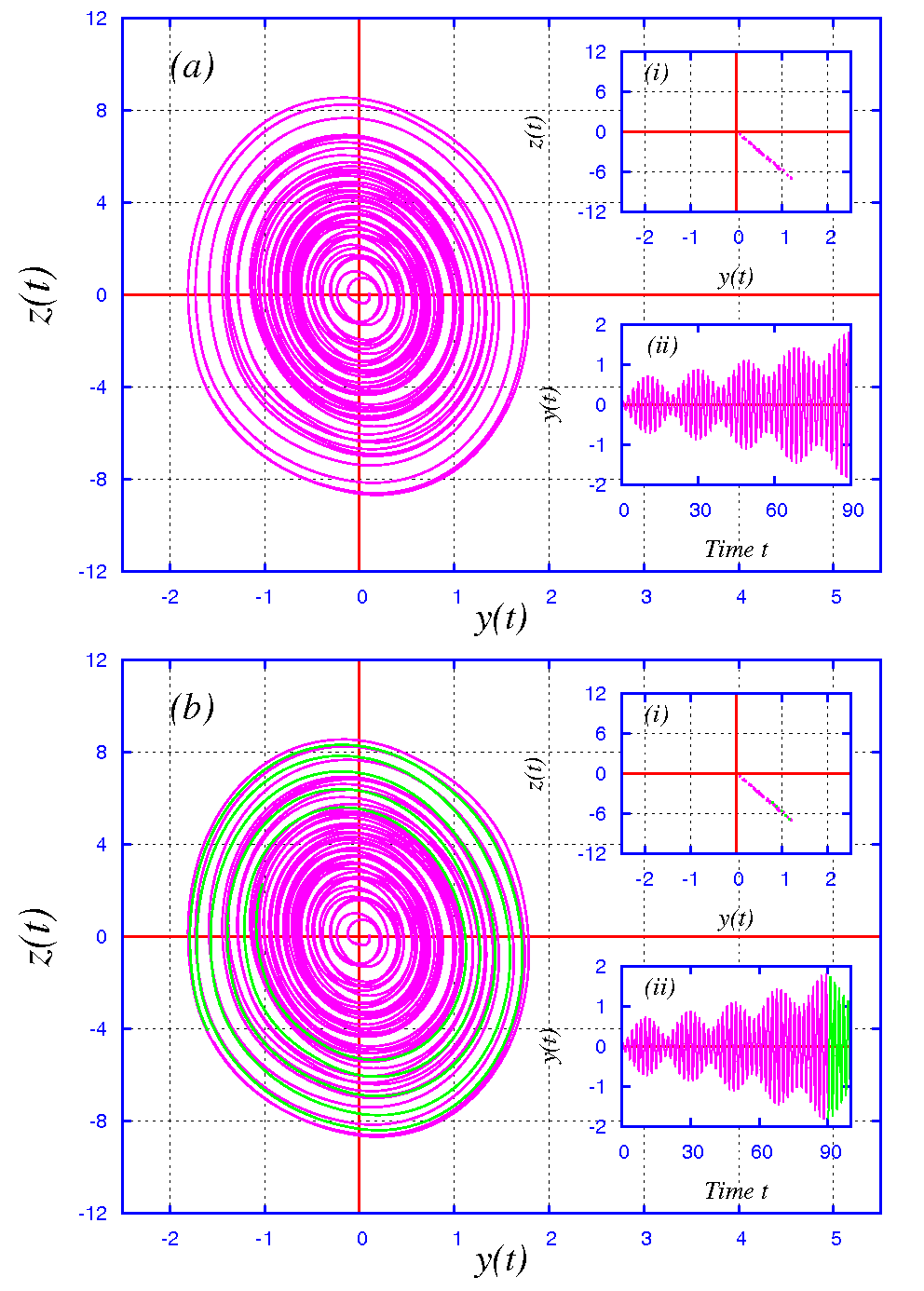}
\caption{Phase portraits of the dynamical variables (numerical analysis) in the normalized voltage $y(t)$ versus normalized current $z(t)$ plane representing (a) the growth of the variables upto 90 time units (online colour –pink), (b) the shrinkage of the variables from 90 time units to 100 time units (online colour –light green), (c) the growth of the variables from 100 time units to 115 time units (online colour –yellow), (d) the shrinkage of the variables from 115 time units to 125 time units (online colour –dark blue) and (e) the growth and shrinkage of the variables from 125 time units to 150 time units (online colour –red). In each figure the corresponding Poincar\'{e} map is depicted as inset (i), while the time series plot representing the growth or shrinkage of the variable $y(t)$ is represented as inset (ii).}
\end{figure}
\addtocounter{figure}{-1}
\begin{figure}
\centering
\includegraphics[width=0.80\columnwidth]{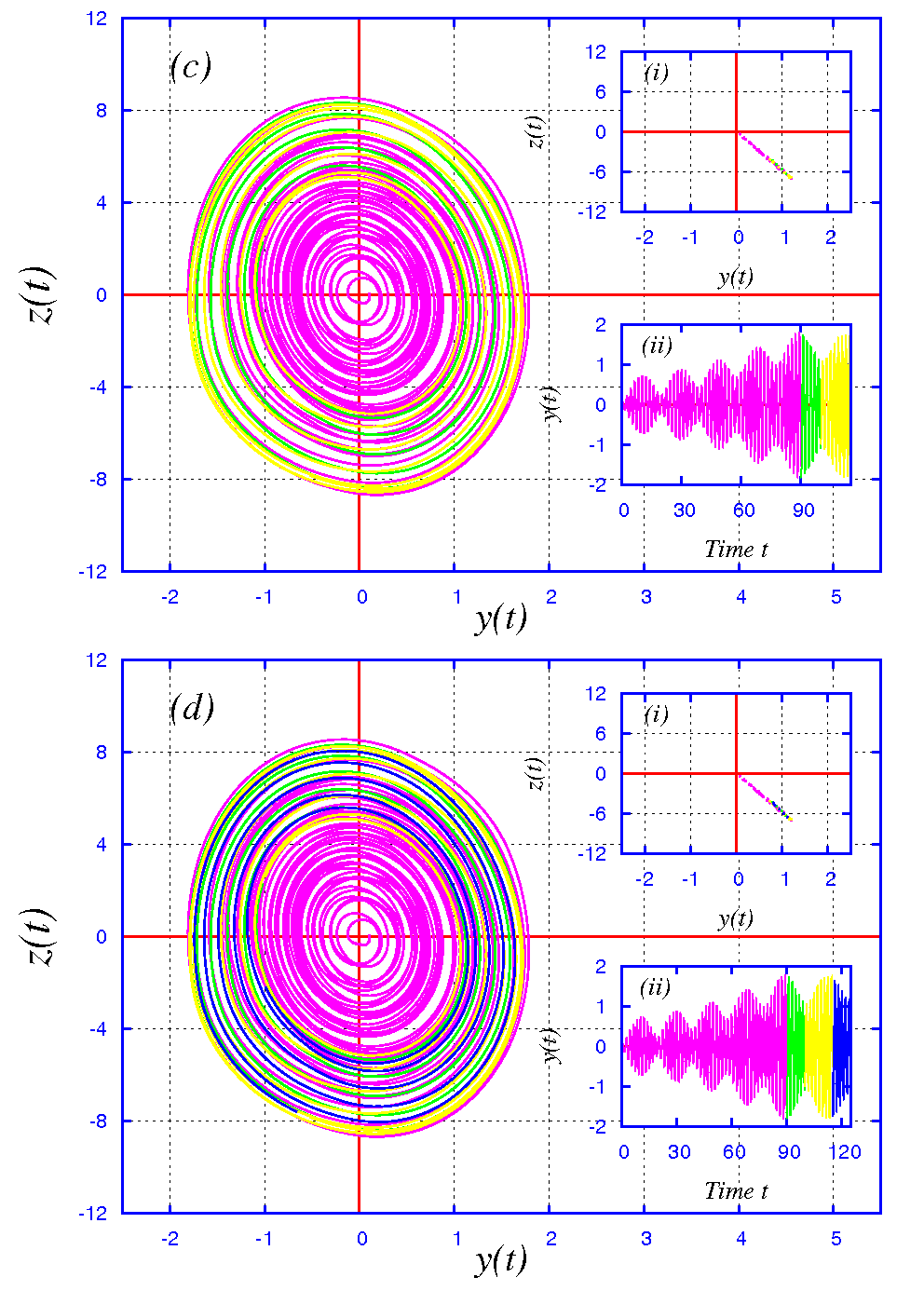}
\caption{(Continued)}
\end{figure}
\addtocounter{figure}{-1}
\begin{figure}
\centering
\includegraphics[width=0.80\columnwidth]{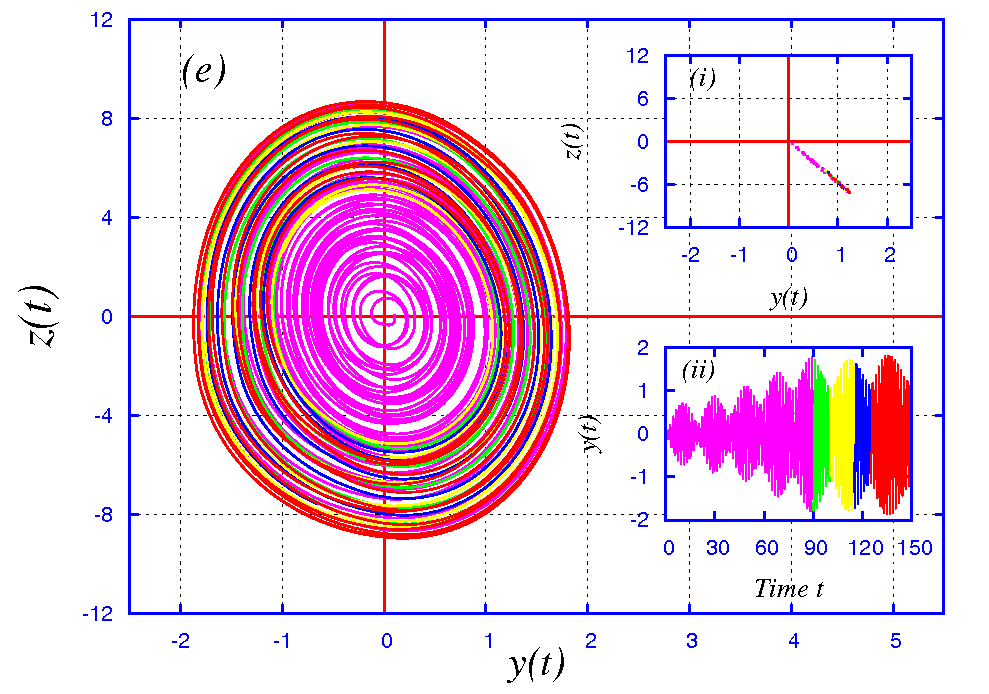}
\caption{(Continued)}
\end{figure}

The power spectrum in Fig. 4(a), is that of the sinusoidal forcing. Evidently it shows a single peak corresponding to the driving frequency $\omega =   
 \omega_{ext} = 3.449394 ~~~(\simeq 3.5)$. The Chua's circuit, by virtue of being a nonlinear oscillatory circuit, generates chaotic self oscillations at a frequency determined by the dynamics of the circuit, even in the absence of a driving force. Hence the power spectrum of the memristor switching, seen in Fig. 4(b), shows a broad band spectrum with a fundamental frequency component. It is this fundamental frequency that determines the switching of the memristor between its memductive states and is identified as the characteristic memristor frequency, $\omega^{*}_{mem} = 3.240534 $. When an external driving force is added, the dynamics of the circuit changes as result of the interaction of the self oscillations and the driving signal. As is expected this interaction results in some sort of control or suppression of chaos that causes the memristor to lower its switching frequency to a value less than its characteristic frequency. This altered memristor frequency component which we call as the new memristor frequency $\omega'_{mem}$ has been identified from the power spectrum in Fig. 4(c) as $\omega'_{mem} = 2.972088 $. Also the driving frequency is identified from the power spectrum as $\omega_{ext} = 3.449394$ as it has to be. The central frequency of the chaotically modulated signal is obviously the average of the memristor and external driving frequencies, namely $\omega_c = (\omega'_{mem} + \omega_{ext})/2$.  From the Fig. 4(c), this central frequency is identified as  $\omega_{c} = 3.240534 $. The power spectrum of the envelope of the chaotically modulated  {\it {y(t)}}-variable is shown in Fig. 4(d).  The frequency of this envelope is often called as the {\it{beat frequency}}. For a damped forced oscillator system, it is half the difference between the two frequencies involved, namely the new memristor and external frequencies, that is $\omega_b = (\omega'_{mem} - \omega_{ext})/2 = 0.270179$.
\begin{figure}
\centering
\includegraphics[width=0.80\columnwidth]{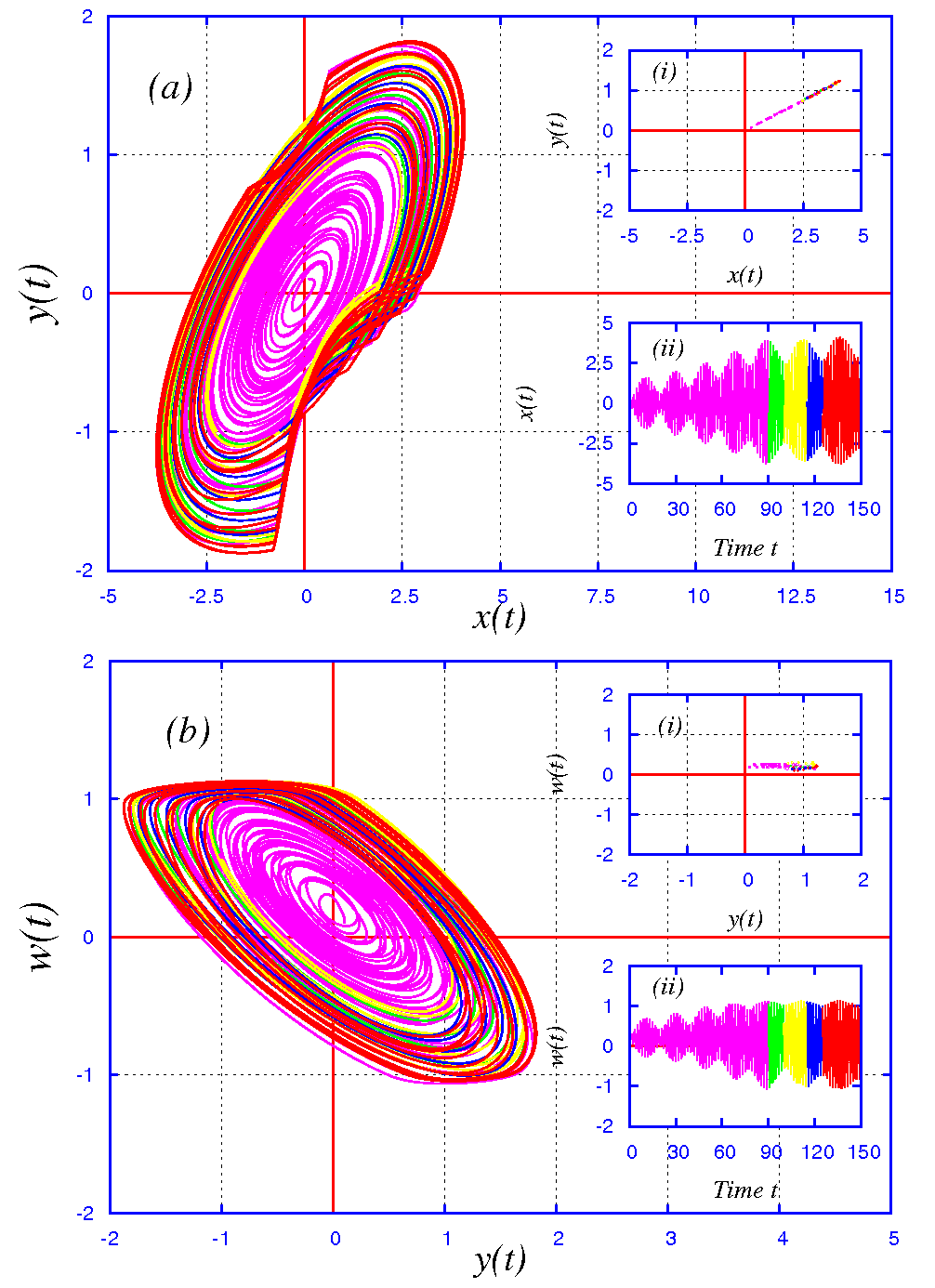}
\caption{Phase portraits (numerical analysis) representing the growth and shrinkage of the dynamical variables from 0 to 150 time units in (a) the normalized voltage $x(t)$ versus normalized voltage $y$ plane, (b) the normalized voltage $y(t)$ versus normalized flux $w$ plane and (c) the normalized current $z$ versus normalized flux $w$ plane. (Online colours are same as those followed earlier). As in the earlier Fig. 5. the corresponding Poincar\'{e} map is depicted as inset (i), while the time series plots representing the growth or shrinkage of the variable $x(t)$, $w(t)$ and $z(t)$ are represented as insets (ii) in each of the phase portraits respectively. }
\end{figure}
\addtocounter{figure}{-1}
\begin{figure}
\centering
\includegraphics[width=0.80\columnwidth]{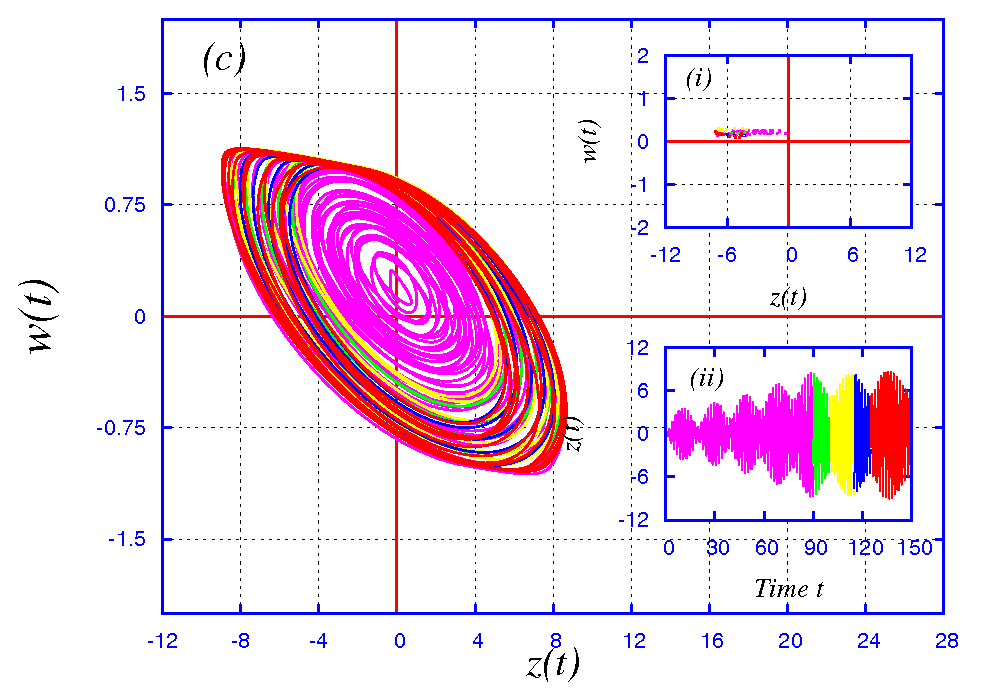}
\caption{(Continued)}
\end{figure}

To confirm the chaotic amplitude modulation, the Lyapunov exponents of the system are obtained using the well known Wolf algorithm [Wolf etal., 1985]. For a forcing amplitude $f = 0.75$ and forcing frequency $\omega = 3.5$, chaotic beats are observed with Lyapunov exponents $( \lambda_1 = 0.000586, \lambda_2= -0.254724, \lambda_3 = -0.264698, \lambda_4 = -21.899207 )$. It is to be noted that the Lyapunov values obtained in this case are much less than that obtained for an autonomous Chua's circuit using a memristor [Itoh $\&$ Chua, 2008], or for a highly frequency detuned driven memristor Chua's circuit wherein beats phenomenon does not arise at all due to the large frequency mismatch. This lowering of Lyapunov exponents denotes a drop in chaoticity of the circuit. This is because of the control of chaos effected by the introduction of the external driving force on the autonomous Chua's circuit. These values, though small, agree well with the Lyapunov exponents obtained in earlier works on Chua's circuits employing ordinary Chua's diode, namely the two coupled autonomous Chua's circuit [Cafagna $\&$ Grassi, 2004] as well as the driven Chua's circuit [Cafagna $\&$ Grassi, 2006a;  2006b] wherein chaotic modulation were observed. While two driving periodic forces were employed to generate beats in the individual circuits, the natural autonomous oscillations of the subsystems of the two coupled autonomous Chua's circuit were used to obtain the same for the latter case. It is pertinent to note here that the present circuit is a single four dimensional non-autonomous circuit (a dimension of two less than the six dimensional non-autonomous coupled Chua's circuit) and generates chaotic beats using a single periodic driving force (unlike two driving forces employed in earlier works). Further the number of dimensionless parameters that tune the behaviour are also less.
\begin{figure}
\centering
\includegraphics[width=1.0\columnwidth]{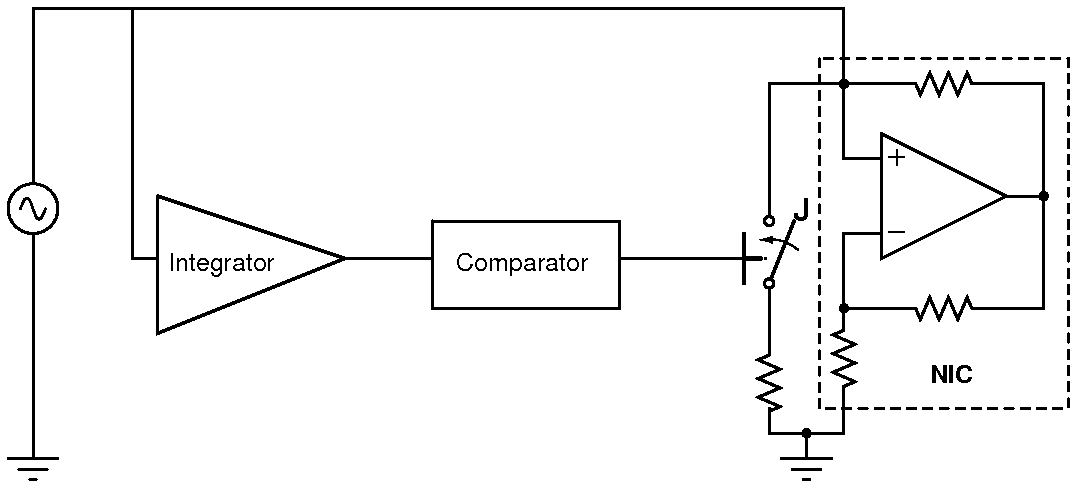}
\caption{(a)The block diagram of the proposed memristor model based on its functioning principle. (b) the actual Multisim model of the memristor used in the study. The values  of the linear resistances are $R_5 = 1200 \Omega$ and $R_6 = 1680 \Omega$. The values of the other parameters are: $R_1 = 10 k\Omega, C_3 = 2.2 nF, R_3 = 50 k\Omega, R_2 = 100 k\Omega, R_4 = 10 k\Omega, R_7 = 2 k\Omega, R_8 = 2 k\Omega$.}
\end{figure}
\addtocounter{figure}{-1}
\begin{figure}
\centering
\includegraphics[width=1.0\columnwidth]{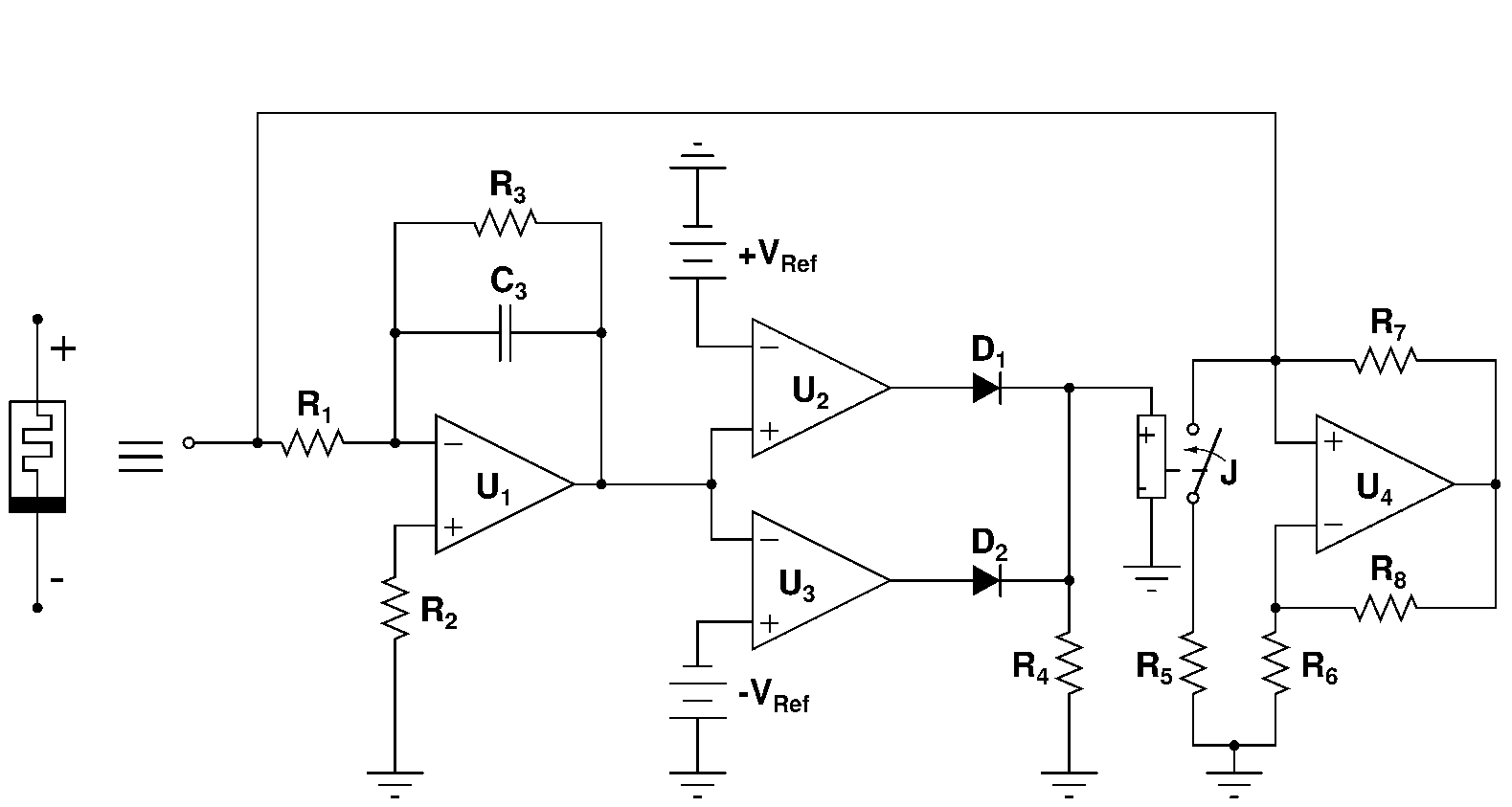}
\caption{(Continued)}
\end{figure}

The phase portraits taken for different time durations of evolution show clearly the growth and the shrinkage in the amplitudes or the revivals and collapses of the system variables as the dynamics evolves, resulting in the occurrence of beats. In Figs. 5$(a-e)$, one finds the successive expansions and contractions of the attractor in the  $(y-z)$ phase plane as time elapses. In Figs. 6$(a-c)$, the corresponding variations in the attractor amplitudes in the  $(x-y)$, $(y-w)$ and $(z-w)$ phase planes are shown. The Poincar\'{e} maps  and the time series plots are shown as insets (i) and (ii) in the respective phase portraits. In all of these cases, the state variables expand up to 90 time units and then undergo a shrinkage up to 100 time units. They undergo a further growth and decay in amplitudes for successive intervals of time. This process of growth and decay in amplitudes continues adinfinitum and is governed by the dynamics of the system. 
\begin{figure}
\centering
\includegraphics[width=0.80\columnwidth]{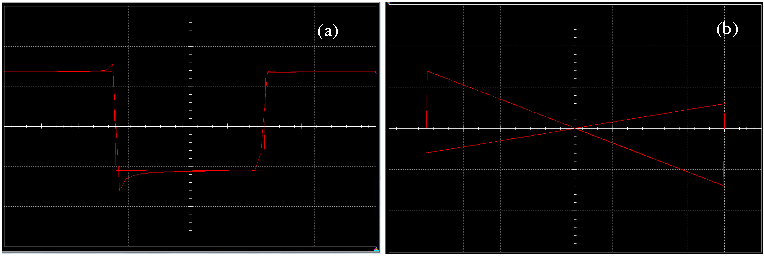}
\caption{(a) The variation of memductance $W(\phi)$ as a function of flux $\phi$ across the memristor (using Multisim model). This is obtained by plotting the output of the integrator versus that of a  divider circuit module whose inputs are the currents and voltages across the memristor. (b) The characterisitc of the memristor in the $(v-i)$ plane. This shows a pinched hysterectic curve, which is a characterisitic feature of memristors. Both of these graphs are obtained from Multisim software grapher facility.}
\end{figure}
\begin{figure}
\centering
\includegraphics[width=0.80\columnwidth]{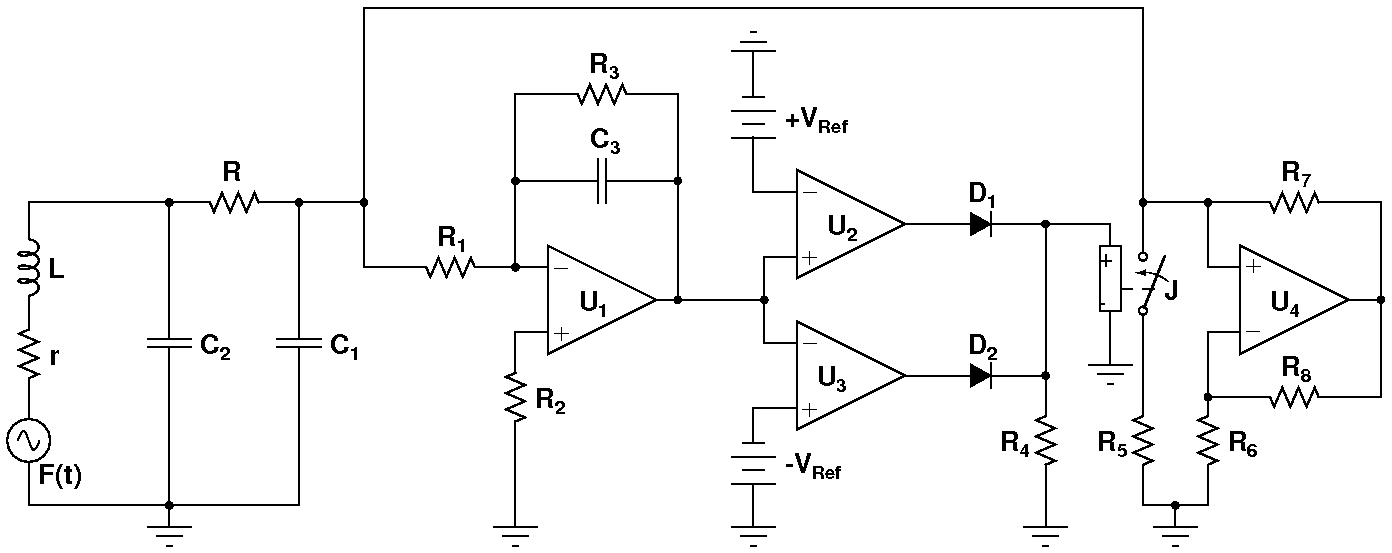}
\caption{The Multisim model of the driven memristor Chua's circuit. The parameters are fixed as $L = 5.95 mH, R = 650 \Omega, C_1 = 5 nF, C_2 = 50 nF,
r = 22.5 \Omega, \nu_{ext} = \Omega/2\pi = 9.555 kHz$ and $F = 740 mV V_{pp}$. The parameters for the memristor are as given earlier.}
\end{figure}
\begin{figure}
\centering
\includegraphics[width=0.8\columnwidth]{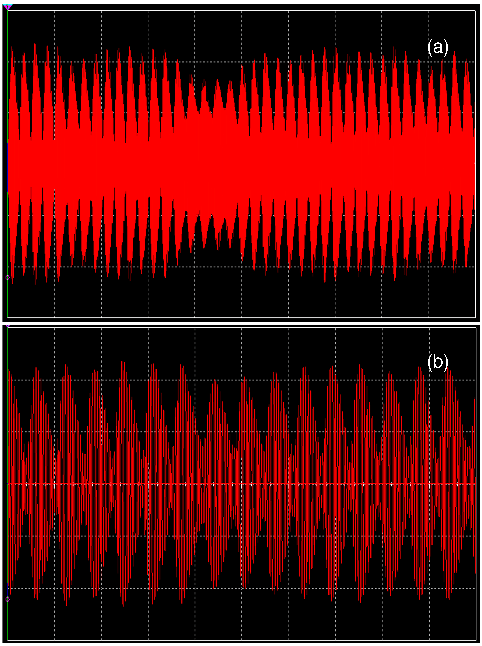}
\caption{Chaotic beats (Multisim modelling):  (a) A section of the time series of the voltage $v_2(t)$ across the capacitor $C_2$ showing the presence of amplitude modulation and chaotic beats. (b) An enlarged portion of the same.}
\end{figure}

\section{\label{Expt}Experimental Investigations}

Next, in this section, we will present our study on the experimental confirmation of the above numerical results through Multisim modelling.
A memristor having a smooth nonlinear third order polynomial relationship between flux and charge has been designed by Muthuswamy [2009b]. This is based on the cubic nonlinear resistive circuit designed by Zhong [1994]. However in this paper, we propose a circuit model for a three segment piecewise flux controlled active memristor. A memristor is a device whose resistive or conductive state depends on the magnitude of flux across it. By Faraday's law, the flux passing through a memristor is the time integral of the electromotive force across it,  

\begin{equation}
\phi = \int_{0}^{t} vdt.
\end{equation}
Using this simple logic, an analog electronic implementation of the memristor can be designed. The block diagram illustrating this principle is given in Fig. 7(a). This model is based the time varying resistor (TVR) circuit proposed by Nishio et al. [1993]. Here a linear resistance and a negative impedance converter (NIC) are switched ON and OFF alternatively based on the output pulse of a comparator.The comparator compares the flux through the memristor ( that is the integral of the voltage across the memristor) between two reference levels, called breakdown points. For flux values lying within the upper and lower breakdown points ( namely $\pm 1$ flux units ) the negative conductance is included in the circuit. For flux values exceeding the breakdown points, the resultant of the linear resistance and negative conductance, which are in parallel, is included in the circuit. The flux is obtained by an integrator circuit. By this action, the functional relationship between the flux and charge given in equation (1) is realized.
\begin{figure}
\centering
\includegraphics[width=0.80\columnwidth]{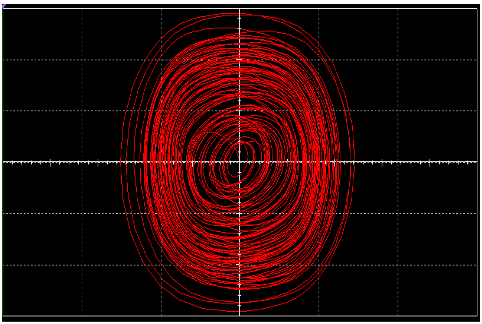}
\caption{Phase portrait (Multisim modelling) representing the growth and shrinkage of the dynamical variables in the $(v_2(t)-\phi(t))$ plane.}
\end{figure}
\begin{figure}
\centering
\includegraphics[width=0.80\columnwidth]{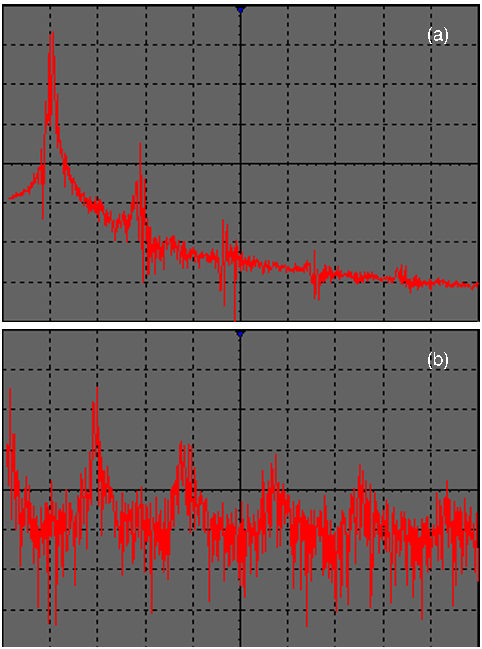}
\caption{The power spectrum (Multisim modelling) of (a) the voltage $v_2(t)$ across the capacitor $C_2$ and (b) the voltage across the memristor. From these the external frequency, the memristor frequency and the central frequency are identified as $\nu_{ext} = 9.6 kHz, \nu'_{mem} = 7.91 kHz$ and $\nu_c = 8.85 kHz$. }
\end{figure}

This proposed circuit is implemented using Multisim modelling and is given in Fig. 7(b). The values  of the linear resistances are $R_5 = 1200 \Omega$ and $R_6 = 1680 \Omega$. The values of the other parameters are: $R_1 = 10 k\Omega, C_3 = 2.2 nF, R_3 = 50 k\Omega, R_2 = 100 k\Omega, R_4 = 10 k\Omega, R_7 = 2 k\Omega, R_8 = 2 k\Omega$.For the Multisim switch module, the ON state resistance is set as $(R_{ON} = 1 m\Omega)$ and the OFF state resistance is chosen as $(R_{OFF} = 1 G\Omega)$ and the threshold voltage is set as $V_{threshold} = 1V)$. When an external sinusoidal source is applied, the flux across the memristor changes, causing the memductance state to switch between a low OFF state and a high ON state as shown in Fig. 8(a). The flux is obtained as the output of the integrator circuit and the divider module facility of Multisim. If the inputs to the divider module are the voltages and currents across the memristor, then an inverse measure of the memductance $W(\phi)$ can be obtained. The characteristic of the memristor in the $(v-i)$ plane is shown in Fig. 8(b). It is clearly a pinched hysterectic curve, wherein the currents fall to zero whenever the voltage across the memristor becomes zero. This is a characteristic feature of the memristors.
\begin{figure}
\includegraphics[width=1.0\columnwidth]{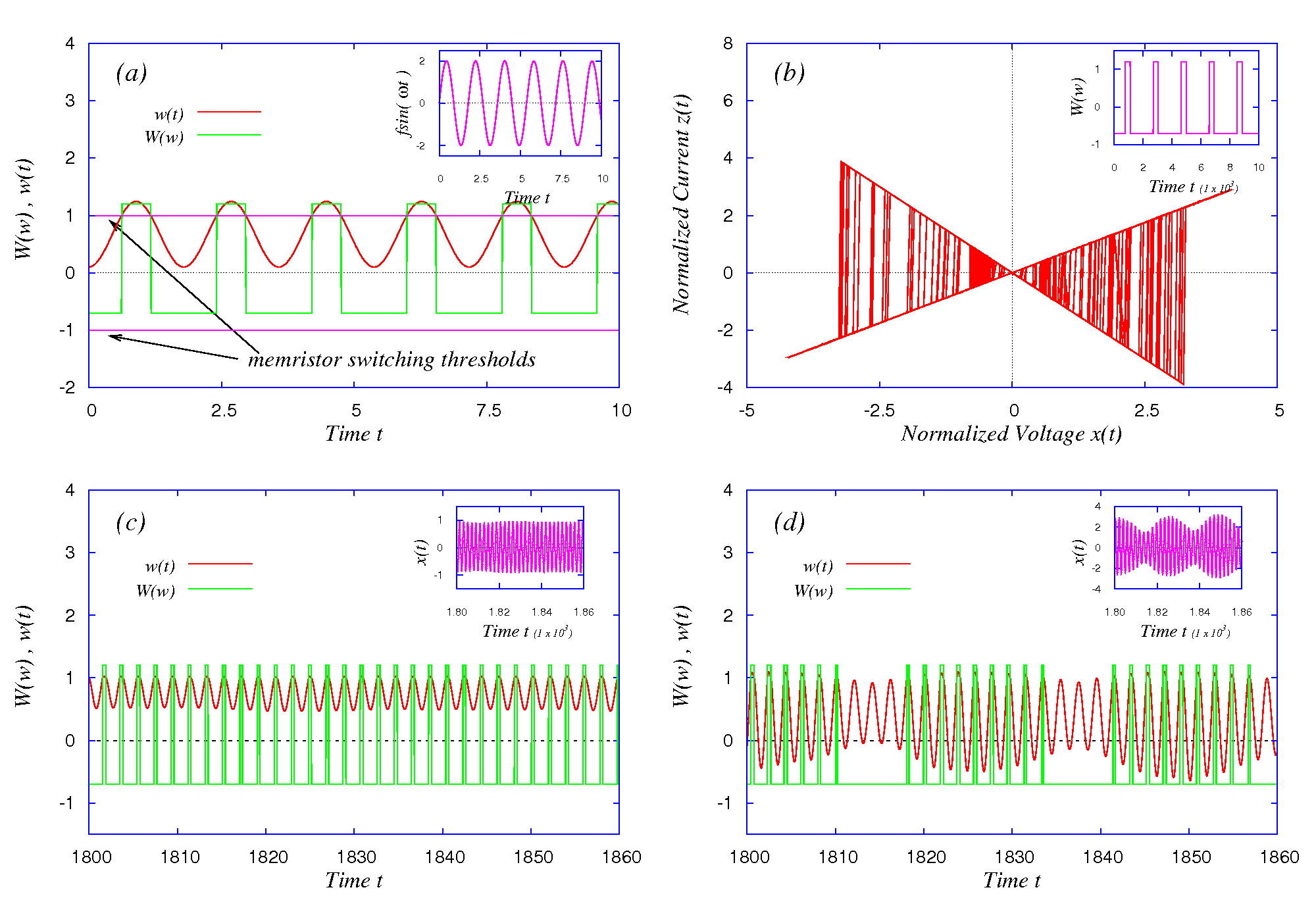}
\caption{Switching characterisitc of the memristor (Numerical): (a) The change in flux arising due to  a linear sinusoidal excitation causing the memristor to toggle between the higher memductance ON state $(b-\xi) = 1.2$ and lower memductance OFF state $(a-\xi) = -0.7$. The threshold levels for flux for this change in memductive states to occur are $\pm 1.0$ flux units. The linear sinusoidal driving signal is shown as inset. (b) The  $v-i$ characteristic of the memristor showing switching between the bi-stable memductance states in the normalized voltage versus current  ($x-z$) plane. The memristor switching as a function of time is shown as inset. (c)The change in flux across the memristor arising due to the self-oscillations of the autonomous Chua's circuit and the consequent memristor switching. The normalized voltage $x(t)$ which drives the memristor is shown as inset. (d) The change in flux across the memristor due to the forced oscillations of the driven Chua's circuit and the consequent memristor switching at an altered frequency. Here also the normalized voltage $x(t)$ which drives the memristor is shown as inset.}
\end{figure}

The driven Chua's circuit is investigated by using Multisim model of the memristor proposed. The circuit implementation is shown in Fig. 9. The parameter values of the circuit  are fixed as $ L = 5.95 mH, R = 650 \Omega, C_1 = 5nF, C_2 = 50nF,r = 22.5 \Omega$. For the memristive part, the parameters are as given earlier. Here, for the sake of experimental convenience, the parameters chosen are slightly different from those of numerical values. Yet the observed behaviours are found to be qualitatively same to those obtained through numerical simulations. For this choice of parameters, chaotic beats are observed when the forcing amplitude is varied keeping the frequency of the external forcing a constant or vice versa. The time series of the voltage across the capacitor $C_2$ showing chaotic variations in amplitude is given in Fig. 10(a). An extended region of the same is given in Fig. 10(b). The phase portrait in the $(v_2(t)-\phi(t))$ is shown in Fig. 11. To obtain these, the frequency of the external sinusoidal forcing is fixed as $\nu_{ext} =  \Omega/2\pi = 9.555 kHz$ and the amplitude is fixed as $F = 740 mV V_{pp}$. The power spectrum of the variable $v_2(t)$ and that of voltage across the memristor $v_1(t)$ are shown in Figs. 12(a $\&$ b) respectively. From these the external frequency, the memristor frequency and the central frequency are identified as $\nu_{ext} = 9.6 kHz$, $\nu'_{mem} = 7.91 kHz$ and $\nu_c = 8.85 kHz$. Note that, this central frequency $\nu_c \simeq (\nu_{ext} + \nu'_{mem})/2$, as expected.
\begin{figure}
\centering
\includegraphics[width=0.8\columnwidth]{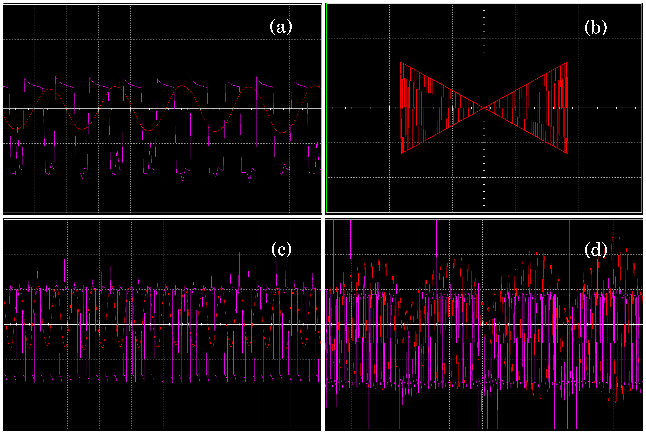}
\caption{Switching characterisitc of the memristor obtained from Multisim modelling: (a) The change in flux arising due to  a linear sinusoidal excitation causing the memristor to toggle between the higher memductance ON state and the lower memductance OFF state. (b) The  $v-i$ characteristic of the memristor showing switching between the bi-stable memductance states. (c)The change in flux across the memristor for parameter values far away from those causing beats and the consequent memristor switching. (d) The change in flux across the memristor at the incidence of beats and the consequent memristor switching.}
\end{figure}

\section{\label{Mech}Mechanism of Chaotic Beats in the Driven Memristive Chua's Circuit }

The mechanism for the generation of beats is obviously the switching of the memristor. Essentially a memristor is a device that works under alternating current (a.c) conditions wherein the applied voltage varies sinusoidally with time. As the polarity of this voltage changes sinusoidally with time, the flux across the memristor also changes, causing it to switch reversibly between a less conductive OFF state (low memductance) and a more conductive ON state (high memductance) [Tour $\&$ He, 2008]. This is shown in Fig. 13(a). The lower memductance level $(a-\xi )$ is the OFF state and the higher level $(b- \xi )$ is the ON state. The  $v-i$ characteristic of the memristor (in the normalized variables)  shown in Fig.13(b), makes one perceive vividly the transitions between the  bi-stable memductive  states. The memristor just acts as a linear time varying resistor (LTVR), see for example[Chua {\it{et al.,}} 1987; Nishio $\&$ Mori, 1993]. This switching of the memristor takes place at a characteristic time period or frequency. However when the memristor is driven by chaotic signals, either through a self oscillatory mechanism or through an external generator of chaotic signals, it switches chaotically between its memductive states. The memristor now acts as a chaotically time varying resistor (CTVR). This chaotic switching is shown in Figs. 13(c) $\&$ 13(d), where the memristor is driven by voltage across the capacitance $C_1$ of the autonomous Chua's circuit or the driven Chua's circuit, for both of which it is an integral part as their nonlinear element. The switching of the memristor observed from Multisim model are qualitative equivalent to Figs 13(a-d) and are shown in Figs. 14(a-d). The unwanted glitches seen in Figs. 14(a $\&$ d), arise due to the mismatch in the delay times of the circuit components.
\begin{figure}
\includegraphics[width=0.8\columnwidth]{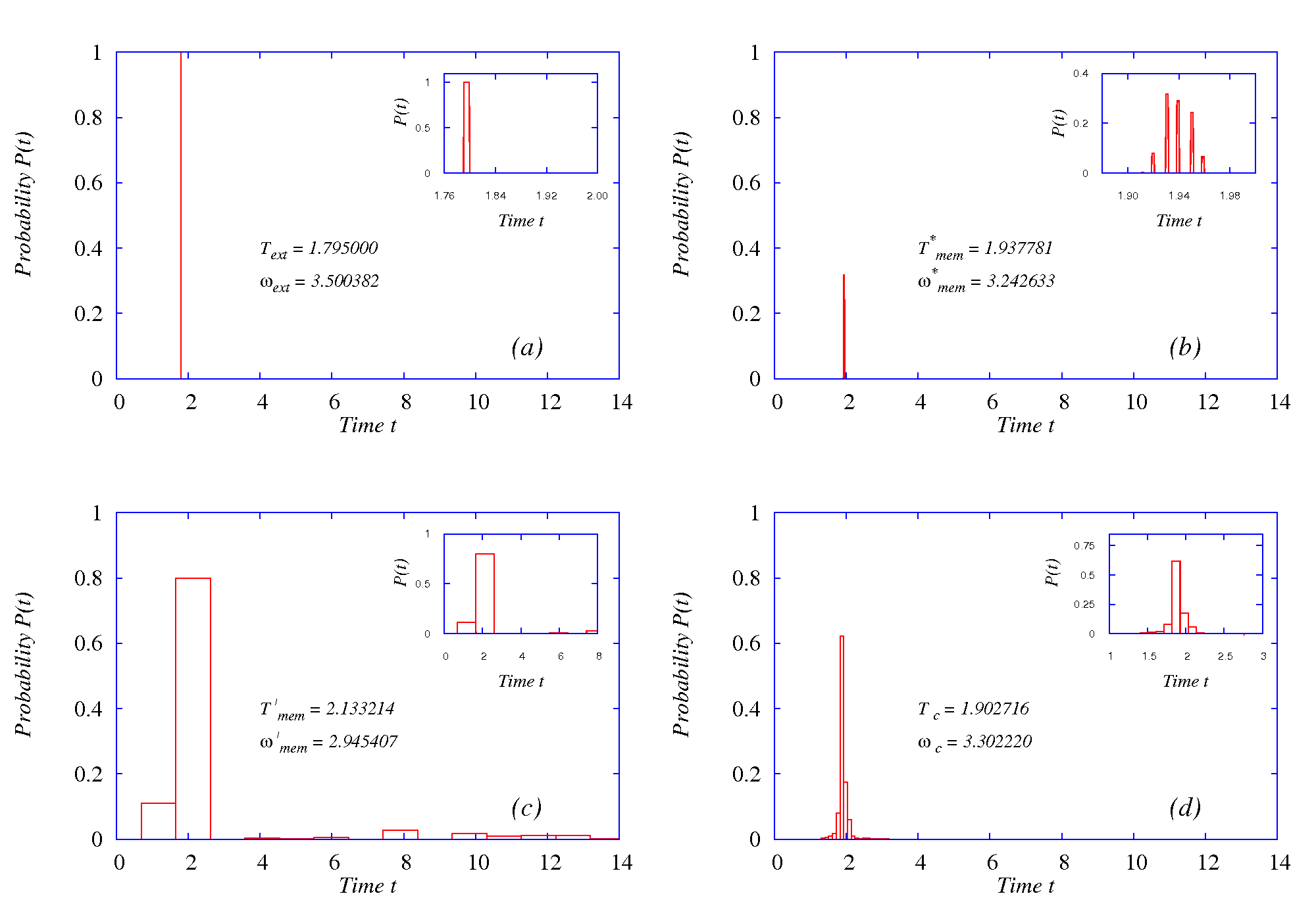}
\caption{Probability of the  switching times (from numerical data) for (a) the (positive and negative cycles) of the driving sinusoidal signal, (b) self oscillations forced memristor switching, (c) driven oscillations forced memristor switching and (d) the (positive and negative cycles) of the variable $y(t)$ of the driven memristive Chua's circuit. The corresponding frequencies are calculated as reciprocals of these switching times. Insets are blown up portions.}
\end{figure}

From the normalized Eqs.(6) $\&$ (7), for the amplitude of the driving force $f=0$, the equilibrium state of the system is given by 
\begin{equation}
A = \{ (x,y,z,w) | x = y = z = 0, w =constant\},
\end{equation}
which corresponds to the $w$-axis. The Jacobian matrix $D$ at this equilibrium state is given by

\begin{equation}
D =   \left ( \begin{array}{cccc}
				-\alpha (W(w)+1) & \alpha & 0 & 0 \\
				1    & -1   & 1   & 0 \\
				0    & -\beta    & -\gamma    & 0 \\
				1    & 0    & 0    & 0 
				\end{array}
		\right)
\end{equation}
Then the characteristic equation is
\begin{equation}
\rho^4 + a_2\rho^3 + a_1\rho^2 +a_0\rho  = 0,
\end{equation}
where the  ${\it{a_i}}$'s,~$i=1,2$ are various coefficients and {\it{$\rho$}}'s are the eigenvalues. Equation (10) can be factorized as
\begin{equation}
\rho(\rho^3 +a_2\rho^2 +a_1\rho^1 +a_0 ) = 0
\end{equation}
Then the natural frequency of the autonomous circuit for $(a_2 = 0 = a_0)$  is evaluated by the relation
\begin{equation}
\omega = \sqrt{a_1}, a_1 = [ ( \beta +\gamma) - \alpha(1+W(w))*(\gamma+1)-\alpha]
\end{equation}
\begin{equation}
W(w)   = \left \{\begin{array}{ll}
				(a-\xi),   ~~~ | w | \leq 1\\
				(b-\xi),   ~~~ | w | > 1				
				\end{array}
		\right. 	
		a,b,\xi > 0
\end{equation}

Numerically, from Eqs. (13) -(14), the frequency of the autonomous circuit is found to vary between $\omega_1 =3.146427 $ and $\omega_2 = 5.962382$, depending on the values of $a$, $b$ and $\xi$.

The breakpoints in the $\phi-q$ characterisitc of the memristor shown in Fig. 2(b), namely $\pm 1$ flux units, can be considered as the memristor switching thresholds. When the flux across the memristor exceeds these threshold values, switching of the memristor between its memductive ON and OFF states occurs. Let $T_a$ and $T_b$ refer to the durations of time the memristor resides in the ON state and OFF state, before switching, respectively. Then the total time taken by the memristor to complete one full  switching process is $T = T_a + T_b$. The switching frequency $ \omega_{mem}$ of the memristor is then 2$\pi$ times the reciprocal $1/T $ of this total switching time. If the memristor switching frequency $ \omega_{mem}$ is  equal to that of the external driving frequency $\omega_{ext}$, then the memristor switching operation is said to be $\it{synchronous}$ or $\it{harmonic}$. If it is double or multiples of the external frequency, then the memristor switching operation is said to be $\it{subharmonic}$.
\begin{figure}
\includegraphics[width=0.8\columnwidth]{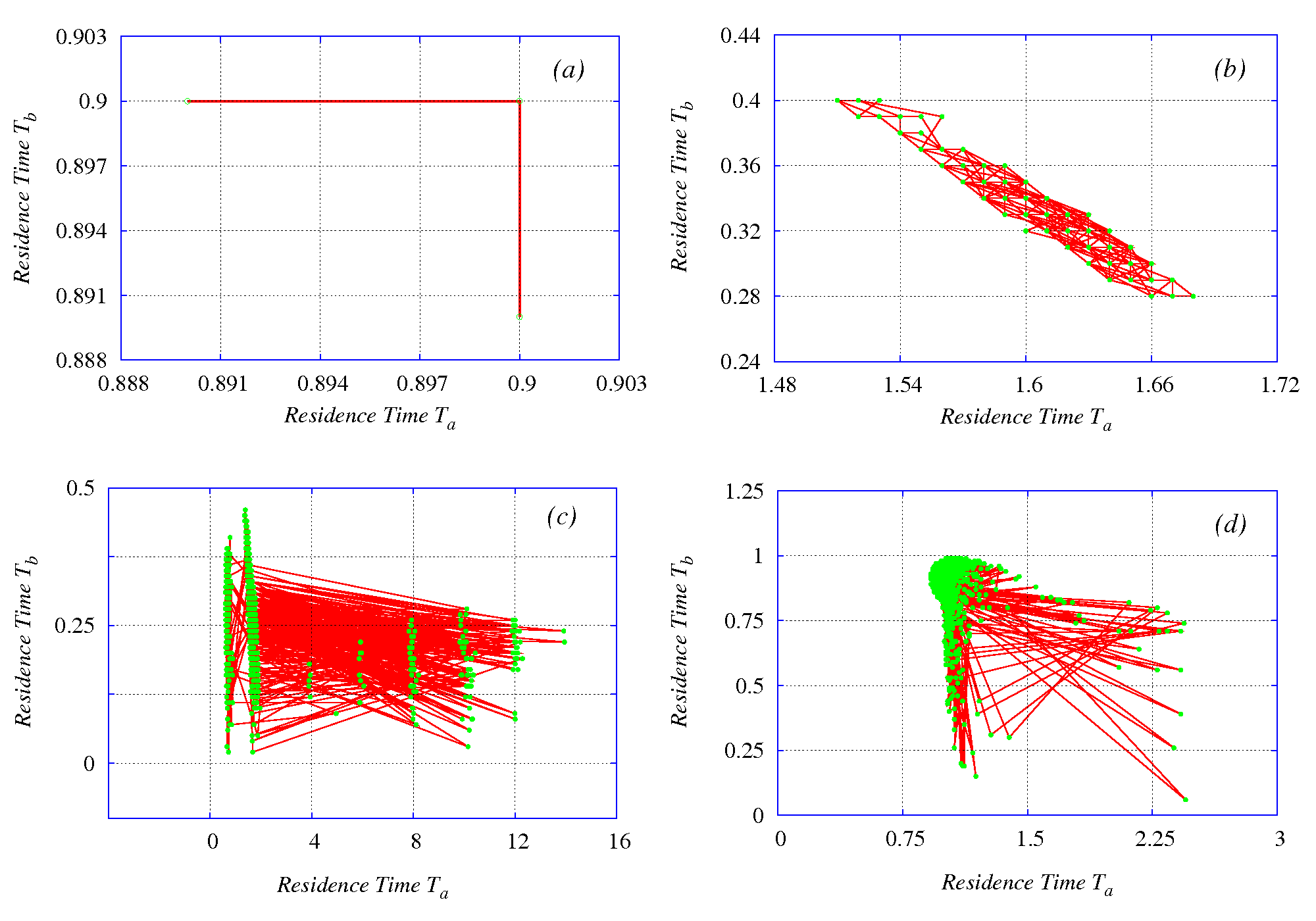}
\caption{Two dimensional plots (numerical analysis) of residence times for the memductive ON and OFF states for (a) linear excitation, (b) self oscillations, (c) forced oscillations and (d) the variable $y(t)$ of the forced oscillations of the driven Chua's circuit.}
\end{figure}

Let us consider a linear sinusoidal excitation of frequency $\omega = 3.5 $ is applied to the memristor. If the flux across the memristor arising due to this sinusoidal excitation were to exceed the two switching thresholds, $\pm 1 $ flux units, then for each external driving cycle, the memristor completes two full switching cycles. The memristor operation will then be subharmonic with a frequency equal to twice the external driving frequency, that is $\omega_{mem} = 2\omega_{ext}$. However, in the present case the flux across the memristor is the time integral of the voltage across it, that is, $\phi_{mem} = \int fsin\omega tdt = (f/\omega)cos\omega t$. This cosinusoidal variation of flux causes it to exceed only one of the two memristor switching thresholds, namely +1 flux unit. Consequently only one memristor switching occurs for each cycle of the sinusoidal forcing. Hence the memristor switching operation is {\it{synchronous}}. If we plot the probability of the switching times of the memristor driven by this linear sinusoidal forcing, we find a single probability peak which corresponds closely to the period of the linear excitation, namely $T_{ext} = 1.793564 $. This translates to a frequency of $\omega_{ext} = 3.503184 $. This is shown in Fig. 15(a). If the memristor is connected in parallel to the capacitance $C_1$ of the autonomous Chua's circuit, the memristor gets driven by the $1/\alpha\beta$ self oscillatory combination of the circuit, with a frequency varying from $\omega_1$ and $\omega_2$. The switching of the memristor, because of the inherent chaotic nature of self oscillations of the autonomous circuit, becomes haphazard proving that the memristor now acts as a chaotically time varying resistor CTVR. We get a number of closely spaced switching time probability peaks as shown in Fig. 15(b). The weighted average of these probability peaks gives a switching time $T^*_{mem}  = 1.937781$  which translates to a switching frequency of $\omega^*_{mem} = 3.242463$.

When an external driving force  of a frequency, let us say $\omega = 3.5$, is added to the circuit, the memristor is driven by the forced oscillations of the circuit. As the forced oscillations are also chaotic, the memristor switching is also necessarily chaotic. Hence it still acts as a CTVR. The probability graph of the memristor switching times shown in Fig. 15(c) now shows a prominent peak at a switching time $T^{\prime}_{mem}  = 2.133214$ which corresponds to a memristor switching frequency of $\omega^{\prime}_{mem} = 2.945407$. This is the new switching frequency that the memristor takes on as a result of the apparent control of chaos effected in the circuit by the introduction of the external force. At these particular frequencies, $\omega_{ext}= 3.503184$ and $\omega^{\prime}_{mem} = 2.945407$, the external excitations and memristive switching aided forced oscillations in the circuit undergo constructive and destructive interferences, resulting in the  modulation of amplitudes and frequencies which we call beats.  Varying the external frequency $\omega_{ext}$ away from this chosen value disrupts the interferences causing the modulation phenomenon to wane and eventually disappear. The average of the external driving frequency and the new memristor frequency gives the central frequency of modulated signal. If we plot the switching time probabilities for the positive and negative cycle variations of the variable $y(t)$ under amplitude modulation, we get a prominent probability peak at a switching time $T_c  = 1.902716$. This corresponds to the central frequency $\omega_{c} = 3.302220$. These are shown in Fig. 15(d). The beat frequency, calculated as half the difference in the above said two frequencies, is $\omega_{b} = 0.278888$. The central and beat frequencies obtained from power spectral calculations as well as from probabilistic considerations are shown in Table.1. Both of these agree well qualitatively with the theoretical results. However the small differences in their numerical values may be explained as due to the chaotic nature of the system as well as the large number of frequencies involved.

The two dimensional residence time  plots for the memristor for the ON and OFF memductive states are plotted in Fig. 16. When a linear force is driving the memristor, the switching (between positive and negative half cycles of the sinusoidal force) is regular as seen in Fig. 16(a). In this case there are only two different values for $T_a$ and $T_b$. When the oscillatory part of the autonomous Chua's circuit drives it, because of the inherent chaotic nature of the circuit, the memristor also switches irregularly. The scatter plot of residence times $T_a$ and $T_b$ is erratic and random. This is evident in Fig. 16(b). When a driving force is added, both the external oscillations and the driven oscillations of the circuit interact with each other. In this interaction a control of chaos in the circuit is effected by the external driving force. As the chaoticity is reduced, the circuit tends to approach a regular behaviour, resulting in the occurrence of beats. However as the dynamics of the circuit is still chaotic, the plot of the memristor switching times is also still chaotic, but with a lesser degree of chaoticity or irregularity. This is also the reason for the low values of the Lyapunov exponents obtained. This  weak chaotic switching is illustrated in Fig. 16(c). The scatter plot of the switching times (between positive and the negative cyclic variations) of the variable $y(t)$ of the system is shown in Fig. 16(d).

\section{\label{Con}Conclusion}

In this paper the presence of chaotic beats phenomenon in a driven memristive Chua's circuit is reported. This phenomenon has been observed and verified using numerical simulations in the form of time series plots, phase portraits and Poincar\'{e} maps. The chaotic nature of the modulation in amplitudes has been characterized with power spectral density calculations and Lyapunov exponents. Using Multisim modelling an analog electronic circuit to realise the action of a three segment flux controlled memristor is proposed, and the presence of chaotic beats in the driven memristor Chua's circuit using this memristor has been verified. Further, based on probabilistic considerations, the mechanism for the occurrence of beats has been found to be the interaction due to the dynamics based chaotic time varying resistive property  and chaotic switching of the memristor between different memristive states. The use of memristor in the driven Chua's circuit has enhanced its dynamics considerably. Many features like reverse period doubling route, intermittent route and higher dimensional torus break down route to chaos, hyper chaos, period adding and Farey sequences, transient chaos, double hook attractors etc., can also be identified in this circuit. Much progress has been made along these lines and the results will be published seperately.

\section{\label{Ack}Acknowledgement}

This work forms a part of a Department of Science and Technology (DST), Government of India, IRHPA project of ML and a DST Ramanna Fellowship awarded to him. A.I acknowledges gratefully the University Grants Comission (UGC) of the Government of India for supporting his work under a FDP fellowship (F.ETFTNBD011 FDP/UGC-SERO).

\newpage

\section*{References}

\begin{description}

\item[]
Anishchenko, V. S., Safonova,M.A.~$\&$~ Chua,L.O.[1993] ~``Stochastic resonance in the nonautonomous 
       Chua's circuit," {\it{ J. Cir. Syst. Comput.}} {\bf{3}}, 553-578.
\item[]
Cafagna, D. ~$\&$~Grassi, G. [2004] ~``A new phenomenon in nonautonomous Chua's circuits: Generation of 
      chaotic beats,"{\it{ Int. J. Bifurcation and Chaos}} {\bf{14}}, 1773-1788.
\item[]
Cafagna, D. ~$\&$~Grassi, G. [2005] ~``On the generation of chaotic beats in simple nonautonomous circuits," {\it{Int. J. Bifurcation and Chaos}} {\bf{15}}, 2247-2256.
\item[]
Cafagna, D. ~$\&$~ Grassi, G. [2006a] ~``Generation of chaotic beats in a modified Chua's circuit Part -I:  
       Dynamic Behaviour," {\it{ Nonlinear Dynamics }} {\bf{44}}, 91-99
\item[]
Cafagna, D. ~$\&$~ Grassi, G. [2006b] ~``Generation of chaotic beats in a modified Chua's circuit Part -II: 
       Dynamic Behaviour," {\it{ Nonlinear Dynamics }} {\bf{44}}, 101-108
\item[]
Chua, L.O. [1971] ~``Memristor- The missing circuit element,"{\it{ IEEE Trans. Circuit Th.  CT-} V }{\bf{18}}, 507-519.

\item[]
Chua, L. O., ~$\&$~ Kang, S. M. [1976] ~``Memristive devices  and systems," {\it{Proc. IEEE}} {\bf{64}}, 209–223.

\item[]
Chua, L.O., ~Desoer, C. A. ~$\&$~ Kuh, E. S. [1987] ~{\it{ Linear and Nonlinear Circuits}} (McGraw-Hill Book Company, Singapore) Chap. 2,pp. 59-61.

\item[]
Elwakil, A. S. [2002]  ~``Nonautonomous pulse-driven chaotic oscillator based on Chua's circuit," {\it{Microelectron.J }} {\bf{33}}, 479-486.
\item[]
Grygiel, K. ~$\&$~ Szlachetka, P. [2002] ~``Generation of chaotic beats,"{\it{ Int. J. Bifurcation and Chaos}} {\bf{12}}, 635-644.

\item[]
Itoh, M. ~$\&$~ Chua, L. O. [2008] ~``Memristor oscillators,"{\it{ Int. J. Bifurcation and Chaos}} {\bf{18}}, 3183-3206.

\item[]
Liu, Z. [2001] ~``Strange nonchaotic attractors from periodically excited Chua's circuit," {\it{ Int. J. Bifurcation  and    
         Chaos}} {\bf{11}}, 225-230.

\item[]
Murali, K. ~$\&$~  Lakshmanan, M. [1990] ~``Observation of many bifurcation sequences in a driven piecewise-linear 
       Circuit," {\it{Phys. Lett. A}} {\bf{151}}, 412-419.

\item[]
Murali, K. ~$\&$~ Lakshmanan, M. [1991] ~``Bifurcation and chaos of the sinusoidally-driven Chua's circuit," {\it{ Int. J. 
       Bifurcation and Chaos}} {\bf{1}}, 369-384.

\item[]
Murali, K. ~$\&$~ Lakshmanan, M. [1992a]~``Transition from quasiperiodicity to chaos and devil's staircase 
      structures  of the driven Chua's circuit," {\it{ Int. J. Bifurcation and Chaos  }} {\bf{2}}, 621-632.

\item[]
Murali, K. ~$\&$~ Lakshmanan, M. [1992b] ~``Effect of sinusoidal excitation on the Chua's circuit,"{\it{IEEE Trans. 
      Circuits \& Syst.-1 }} {\bf{39}}, 264-270.

\item[]
Murali, K. ~$\&$~ Lakshmanan, M. [1993] ~``Chaotic dynamics of the driven Chua's circuit," {\it{IEEE Trans. Circuits ~$\&$~ 
       Syst.-1}} {\bf{40}}, 836-840.

\item[]
Muthuswamy, B.[2009a] ~``Memristor based chaotic circuits," {\it{ Technical Report No. UCB/EECS- 2009-6}}  
        http://www.eecs.berkely.edu/Pubs/TechRpts/2009/EECS-2009-6.html
\item[]
Muthuswamy, B.[2009b] ~``Implementing Memristor based chaotic circuits,"{to be published in \it{Int. J. 
       Bifurcation and Chaos}}.

\item[]
Nishio, Y. ~$\&$~ Mori, S. [1993] ~``Chaotic phenomena in nonlinear circuits with time-varying resistors," {\it{IEICE Trans. Fundamentals E-76A}} {\bf{3}}, 467-475.

\item[]
\'{S}liwa, I., Grygiel, K. ~$\&$~ Szlachetka, P.[2008] ~``Hyperchaotic beats and their collapse to the quasiperiodic  
        oscillations," {\it{ Nonlinear Dynamics }} {\bf{53}}, 13-18.

\item[]
Strutkov, D.B., Snider, G.S., Stewart, D.R. ~$\&$~ Williams, R.S. [2008] ~``The missing memristor found," {\it{Nature}} {\bf{453}}, 80-83

\item[]
Srinivasan, K., Thamilmaran, K. ~$\&$~ Venkatesan, A. [2009] ~``Classification of bifurcation and chaos in Chua's circuit   
        with effect of different periodic forces," {\it{ Int. J. Bifurcation and Chaos}} {\bf{19}}, 1951-1973.

\item[]
Tour, J. M. ~$\&$~ Tao He. [2008] ~``The fourth element," {\it{Nature}} {\bf{453}}, 42.

\item[]
Wolf, A., Swift,J.B., Swinney, H.L. ~$\&$~ Vastano, J.A. [1985] ~``Determination of Lyapunov exponents from a time 
      series," {\it{ Physica}} { \bf{16D}}, 285-317.
\item[]
Zhong, G. [1994] “Implementation of Chua’s Circuit with a Cubic Nonlinearity,” IEEE Transactions on
    Circuits and Systems 41(12), pp. 934-941.

\item[]
Zhu, Z. ~$\&$~ Liu, Z. [1997] ~``Strange nonchaotic attractors  of Chua's circuit with quasiperiodic excitation," {\it{ Int. J.  
      Bifurcation and Chaos }} {\bf{7}}, 227-238.
                                            
\end{description}

\section*{Table}

\centering 
\begin{threeparttable}
\caption{Comparison of frequencies obtained from power spectral and switching time probability calculations from numerical analysis.\tnote{$\dagger$}}
\begin{tabular} {ccc}\\ \hline\hline \\
\multicolumn{1}{c}{Frequencies}&
\multicolumn{1}{c}{Spectral density calculation}&
\multicolumn{1}{c}{Probabilistic consideration}	\\ \\ \hline\\

External driving frequency \\
$\omega_{ext}$     			&   3.499394			&   3.503184 \\ \\

Memristor switching frequency\\
 $\omega^{*}_{mem}$   			&   3.240534			&   3.242463 \\ \\

Memristor switching frequency\\
  $\omega'_{mem}$ 			&   2.972088			&   2.945407 \\ \\

Central frequency of the modulated \\
signal  $\omega_c = \frac{\omega'_{mem} +\omega_{ext}}{2}$ 
					&   3.240534			&   3.302220  \\ \\
Beat frequency 
 $\omega_b = \frac{\omega'_{mem} -\omega_{ext}}{2}$
					&   0.270179			&   0.280360\\ \\ \hline\hline \\
                                                                 
\end{tabular}
\begin{tablenotes}
\item[$\dagger$]\scriptsize{ The $y(t)$-variable was considered for taking the power spectrum. A total of 65,536 data points at a step size of 0.01 normalized time units were taken for evaluaiton. For the power spectrum of the envelope an average time step was calculated as 1.890656 normalized time units. A total of 4,096 data points of the envelope were taken from a time series of 500,000 data points of the variable $y(t)$ after leaving out transients of 150,000 points. For switching time probabilities a total of k/100 bins for grouping the data were taken, k being the number of times switching occurred in the time interval considered.}
\end{tablenotes}
\end{threeparttable}

\end{document}